\documentclass[twocolumn]{aastex62}
\usepackage{amsmath}

\usepackage{xcolor}

\graphicspath{{./}{figures/}}

\received{June 1, 2019}
\revised{\today}
\accepted{\today}

\submitjournal{ApJ}

\shorttitle{Water Vapor Detection in Earth Analog Reflection Spectra}
\shortauthors{Smith et al.}

\begin{document}

\title{Detecting and Characterizing Water Vapor in the Atmospheres of Earth Analogs \\ through Observation of the $0.94\,\mu$m Feature in Reflected Light}

\author{Adam J.  R.  W.  Smith}
\affiliation{Department of Astronomy \& Astrophysics, University of California, Santa Cruz, CA 95064, USA}

\author{Y.  Katherina Feng}
\affiliation{Department of Astronomy \& Astrophysics, University of California, Santa Cruz, CA 95064, USA}
\affiliation{NSF Graduate Research Fellow}
\affiliation{University of California, Santa Cruz, Other Worlds Laboratory}

\author{Jonathan J.  Fortney}
\affiliation{Department of Astronomy \& Astrophysics, University of California, Santa Cruz, CA 95064, USA}
\affiliation{University of California, Santa Cruz, Other Worlds Laboratory}

\author{Tyler D.  Robinson}
\affiliation{Department of Physics \& Astronomy, Northern Arizona University, Flagstaff, AZ 86011, USA}
\affiliation{NASA Astrobiology Institute's Virtual Planetary Laboratory}

\author{Mark S.  Marley}
\affiliation{NASA Ames Research Center, Moffett Field, CA 94035, USA}

\author{Roxana E.  Lupu}
\affiliation{NASA Ames Research Center, Moffett Field, CA 94035, USA}
\affiliation{Bay Area Environmental Research Institute, Petaluma, CA 94952, USA}

\author{Nikole K.  Lewis}
\affiliation{Department of Astronomy and Carl Sagan Institute, Cornell University, 122 Sciences Drive, Ithaca, NY 14853, USA}



\begin{abstract}

The characterization of rocky, Earth-like planets is an important goal for future large ground- and space-based telescopes.  In support of developing an efficient observational strategy, we have applied Bayesian statistical inference to interpret the albedo spectrum of cloudy true-Earth analogs that include a diverse spread in their atmospheric water vapor mixing ratios.  We focus on detecting water-bearing worlds by characterizing their atmospheric water vapor content via the strong 0.94$\,\mu$m H$_2$O absorption feature, with several observational configurations.  Water vapor is an essential signpost when assessing planetary habitability, and determining its presence is important in vetting whether planets are suitable for hosting life.  We find that R=140 spectroscopy of the absorption feature combined with a same-phase green optical photometric point at $0.525-0.575\,\mu$m is capable of distinguishing worlds with less than $0.1\times$ Earth-like water vapor levels from worlds with $1\times$ Earth-like levels or greater at a signal-to-noise ratio of 5 or better with $2\sigma$ confidence.  This configuration can differentiate between $0.01\times$ and $0.1\times$ Earth-like levels when the signal-to-noise ratio is 10 or better at the same confidence. However, strong constraints on the water vapor mixing ratio remained elusive with this configuration even at signal-to-noise of 15.  We find that adding the same-phase optical photometric point does not significantly help characterize the H$_2$O mixing ratio, but does enable an upper limit on atmospheric ozone levels.  Finally, we find that a 0.94$\,\mu$m photometric point, instead of spectroscopy, combined with the green-optical point, fails to produce meaningful information about atmospheric water content.

\end{abstract}




\section{INTRODUCTION}
\label{sec:intro}

In the decades since the landmark discovery of a planet orbiting another Sunlike star \citep{mayor_queloz_1995}, the field of exoplanetary science has grown tremendously.  Thousands of exoplanets have now been found, and the Transiting Exoplanet Survey Satellite (TESS) is expected to find tens of thousands more \citep{ricker2014transiting, huang2018expected}.  The year 2002 saw the first detection of an atmosphere on an exoplanet \citep{charbonneau_brown_noyes_gilliland_2002} and the field of exoplanetary atmospheres has expanded rapidly since that time \citep{marley2006atmospheres, seager2010exoplanet, crossfield2015observations, kaltenegger2017characterize, madhusudhan2019exoplanetary}.  The future study of exoplanetary atmospheres is of major interest to the astronomical community, with several proposed flagship telescopes set to make exoplanetary atmospheric characterization a major mission objective \citep{mennesson2016habitable, bolcar_aloezos_bly_collins_crooke_dressing_fantano_feinberg_france_gochar_et_al._2017, cooray2017origins}.  


A goal of future exoplanet science and atmospheric studies is the discovery and characterization of an Earth analog.  Such a terrestrial planet would reside in the Habitable Zone of its host star -- the orbital distance where liquid water could exist at the surface of a planet \citep[e.g.,][]{kasting_whitmire_reynolds_1993}.  An attractive pathway to characterize such a planet around a Sunlike parent star would be via direct imaging and spectroscopy of light scattered (``reflected") from the planet's surface and atmosphere \citep[e.g.,][]{feng_robinson_fortney_lupu_marley_lewis_macintosh_line_2018}.  As such, two flagship scale missions currently under study, LUVOIR \citep{bolcar_aloezos_bly_collins_crooke_dressing_fantano_feinberg_france_gochar_et_al._2017,roberge2018large} and HabEx \citep{mennesson2016habitable,gaudi2018habitable}, along with a probe-class external occulter to WFIRST \citep{seager2018starshade}, make the detection of reflected light from rocky planet atmospheres a major science goal.


When designing such a telescope, and optimizing a proposed observing strategy, it is valuable to understand what information can be gained from optical photometry and spectroscopy.  In a previous paper \citep{feng_robinson_fortney_lupu_marley_lewis_macintosh_line_2018}, we developed the first retrieval model for terrestrial planet reflection spectra, which built off our previous efforts for optical reflection spectra for giant planets \citep{lupu_marley_lewis_line_traub_zahnle_2016, nayak2017atmospheric}.  \citet{feng_robinson_fortney_lupu_marley_lewis_macintosh_line_2018} investigated optical spectra of the Earth from 0.4 to $1.0\,\mu$m at a range of spectral resolutions to understand a broad range of science questions, including one's ability to constrain the abundances of atmospheric gases, or merely detect their presence.  In addition, we studied potential constraints on planetary radius, cloud parameters, and surface gravity.

In the follow-up investigation presented here, we focus on a specific potential characterization strategy that future large space telescopes may use.  Potentially interesting planets, in or near the habitable zone, will likely be detected by a search of nearby stars via single-band optical photometry.  It is likely that such a detection will be performed at or near the peak brightness of the host star; for a G-type star comparable to our Sun, this peak is roughly $500-600$nm.  After such a planet has been detected, a ``follow the water'' strategy may next ask: Does the detected planet have water vapor in its atmosphere? If so, how much?

In an initial effort to provide quantitative guidance to these questions, here we build on the work of \citet{feng_robinson_fortney_lupu_marley_lewis_macintosh_line_2018} with an eye towards probing the 0.94$\,\mu$m water absorption band, the strongest at optical wavelengths.  Determining if a planet of interest has water (and how much) would be an important milestone in determining if the planet should be followed up with additional spectroscopy.  The ability to constrain the presence of water vapor in an exoplanet's atmosphere is one useful tool that may be used to guide the search for life on other planets \citep{schwieterman2018exoplanet}.  Following our previous work, and in order to make the problem tractable at this stage, we focus on Earth analog planets, meaning current Earth atmospheric abundances, but now with a water vapor mixing ratio that varies across a factor of a thousand, with atmospheric water vapor from 0.01$\times$ that of Earth, to $10\times$ more.  We investigate the relative ability of photometry and R$\,=140$ spectroscopy, at a variety of signal-to-noise ratios (SNR), across the 0.94$\,\mu$m band, to quantify a detection of atmospheric water vapor and constrain its abundance.  A concentration on a single optical band is motivated by the expectation that future space telescope missions must make multiple observations over limited spectral ranges in order to assemble a spectrum \citep[e.g. LUVOIR;][]{bolcar_aloezos_bly_collins_crooke_dressing_fantano_feinberg_france_gochar_et_al._2017}.




In Section~\ref{sec:methods}, we describe the methods used in this study.  In Section~\ref{sec:results}, we describe the results of the investigation.  In Section~\ref{sec:disc}, we discuss these results and draw conclusions from them in an attempt to answer the above questions.  We also suggest paths for future work.


\section{Methods} \label{sec:methods}


\subsection{Albedo Model and Simulated Data} \label{sec:methods:data}

\begin{deluxetable*}{llll}[!]
    \tablecaption{``Earth-like'' atmosphere parameter set}
    \tablewidth{0pt}
    \tablehead{
    \colhead{Parameter} & \colhead{Description} & \colhead{Value} & \colhead{Prior}
    }
    \startdata
    log(H$_2$O) & Water mixing ratio & log($3\times 10^{-3}$) & [-8, 0] \\ 
    log(O$_2$) & Molecular Oxygen mixing ratio & log(0.21) & * \\
    log(O$_3$) & Ozone mixing ratio & log($7\times 10^{-7}$) & [-10, -1] \\
    log(CH$_4$) & Methane mixing ratio & log($1.8\times 10^{-6}$) & * \\
    log(CO$_2$) & Carbon Dioxide mixing ratio & log($400\times 10^{-6}$) & * \\
    R$_p$ [R$_{\Earth}$] & Planet Radius & 1 & [0.5,12] \\
    log(P$_0$) [bar] & Surface Atmospheric Pressure & log(1) & [-2, 2] \\
    log(g) & Surface Gravity & log(9.8) & [0,2] \\
    log(A$_s$) & Surface Albedo & log(0.05) & [-2,0] \\
    log(p$_t$) [bar] & Cloud Top Pressure & log(0.6) & [-2, 2] \\
    log($\delta$p) (bar) & Cloud Thickness & log(0.1) & [-3, 2] \\
    log($\tau$) & Cloud Optical Depth & log(10) & [-2, 2] \\
    log(f$_c$) & Cloud Coverage & log(0.5) & [-3, 0] \\
    \enddata
    \tablecomments{The base parameters used to represent an ``Earth-like" planet in our forward model, including input values and the range of the prior used during the retrieval process.  Parameters marked with * for their prior were not retrieved, and were instead fixed in the nested sampling program at the values given here. In addition, individual models were run with $0.01\times$, $0.1\times$, $1\times$, and $10\times$ the H$_2$O value described here (see Section~\ref{sec:intro}).}
    \label{table:params}
\end{deluxetable*}

To generate model planetary albedo spectra we employ the high-resolution albedo spectra model described in \citet{marley1999reflected}, which was extensively revised in \citet{cahoy_marley_fortney_2010}.  The model was later paired with an MCMC driver in \citet{lupu_marley_lewis_line_traub_zahnle_2016} and \citet{nayak2017atmospheric} to explore the retrieval of atmospheric parameters for gas giant exoplanets at full phase, and crescent phases, respectively.  The Cahoy et al.  and Lupu et al.  papers have extensive descriptions of the model setup.  More recently, \citet{feng_robinson_fortney_lupu_marley_lewis_macintosh_line_2018} modified the code again to treat the surfaces and atmospheres of Earth-like terrestrial exoplanets.  A fuller description can be found there, as we use the same setup for our work.

The three-dimensional albedo model divides a spherical world into a ``disco-ball" of plane-parallel facets.  For each facet, we calculate $\mu_s$, the angle (relative to the zenith) from which downwelling stellar radiation is incident on the facet.  We also calculate $\mu_o$, the scattering angle (again, relative to the zenith) required for emergent light to reach the observer.  At each facet, the model atmosphere utilizes a fixed pressure level grid, and a radiative transfer calculation is performed to determine the emergent intensity at the required zenith and azimuth angles $(\mu_o,\phi_o)$ of the observer.  We take $I(\tau,\mu,\phi)$ to be the wavelength-dependent intensity at optical depth $\tau$, in the direction described by zenith angle $\mu$ and azimuth angle $\phi$.  Thus, the quantity we wish to find for each facet is $I(\tau=0,\mu_o,\phi_o)$. To compute this value for each facet, we follow the steps laid out in \citet{feng_robinson_fortney_lupu_marley_lewis_macintosh_line_2018}, sections 2.1 and 2.2. We implement a two-term Henyey-Greenstein (TTHG) phase function \citep{kattawar1975three} to treat the directly scattered radiation and Legendre polynomials to represent the azimutally-averaged diffusely scattered radiation. We have since updated the forward and backward scattering portions of the TTHG phase function as presented in \citet{cahoy_marley_fortney_2010}, which were specifically tailored for water clouds, to be consistent with the parameterization described in \citet{kattawar1975three} instead.

With $I(\tau$=0$,\mu_o,\phi_o)$ in hand for each facet, Chebychev-Gauss quadrature \citep{horak1950diffuse,horak1965calculations} is used to integrate the total planetary intensity at a given wavelength.  By repeating this procedure at each wavelength of interest, we are able to build up an albedo spectrum across a given wavelength range.

\begin{figure}[!]
    \includegraphics[width=3.3in]{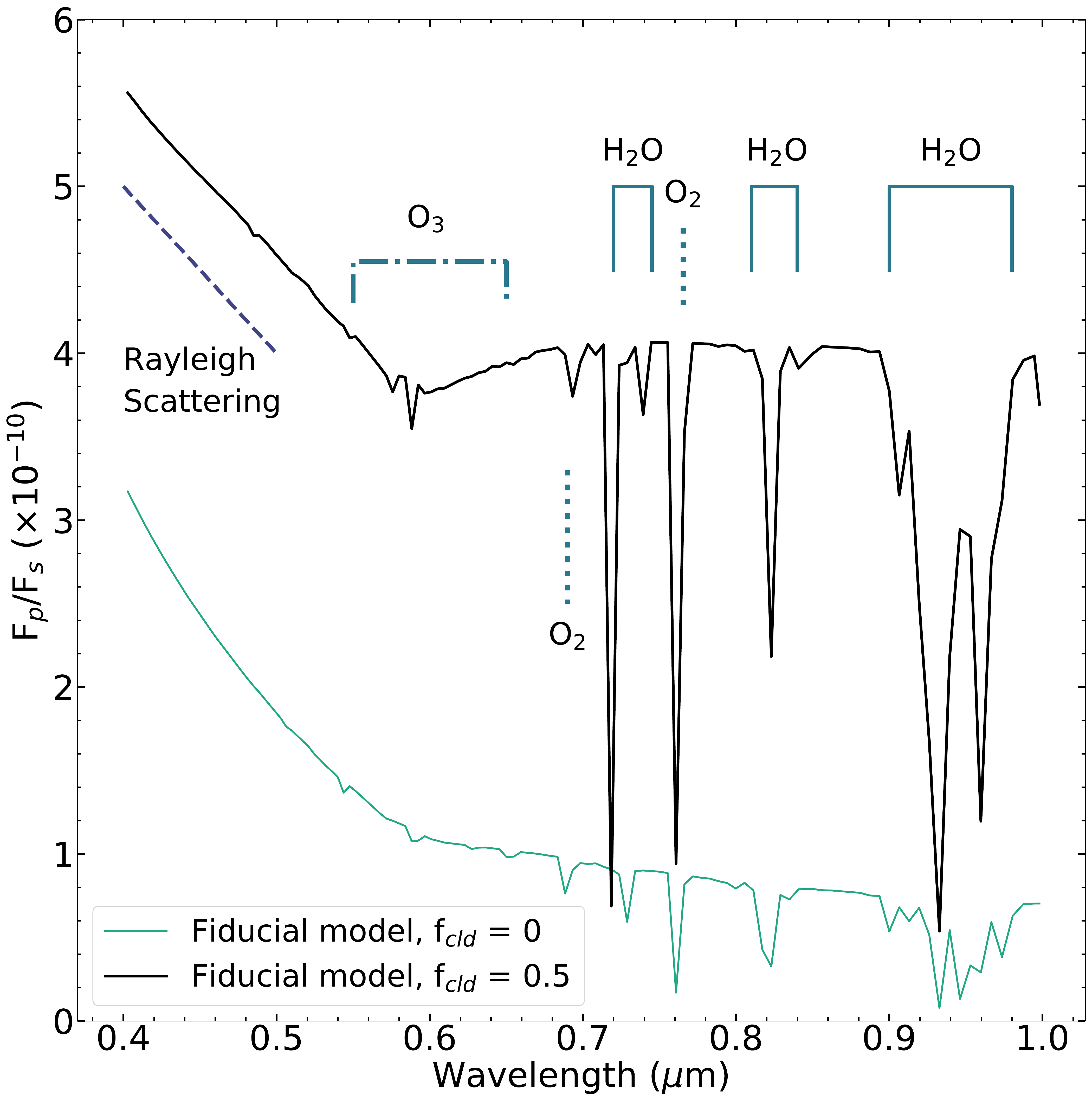}
    \caption {The optical planet-to-star flux ratio spectrum of an Earth-like planet, with a spectral resolution of 140, as generated by the forward model used in this paper, with major features highlighted.  Of interest to this work is the H$_2$O absorption feature at $0.94\,\mu$m.  Note also the lack of major features from non-water sources in the vicinity of this feature.}
    \label{fig:features}
\end{figure}

We focus on the H$_2$O spectral feature centered on 0.94\,$\mu$m. This feature is the strongest one at optical wavelengths where reflected light spectroscopy of potentially habitable planets is most efficient. Stronger features do exist at longer near-IR wavelengths, but there is much less incident flux there from solar type stars and it can be more difficult to obtain spectra due to inner working angle constraints for coronagraphic masks. Notional plans for terrestrial planet characterization in reflected light typically give priority to the detection of this band as an indicator of atmospheric water. We use tabulated H$_2$O opacities --- as well as opacities for O$_2$ and O$_3$ --- generated by the Line-By-Line ABsorption Coefficient model (LBLABC; developed by D.~Crisp; \citet{meadows1996ground}) constructed from the HITRAN~2012 line list \citep{rothman2013hitran2012} \footnote{HITRAN~2012 cites the following references in construction of their water vapor line list: \citet{shirin2006spectroscopically}, \citet{barber2006high}, \citet{brown2007co2}, \citet{furtenbacher2007marvel}, \citet{lisak2008low}, \citet{tennyson2009iupac}, \citet{tennyson2010iupac}, \citet{rothman2010hitemp}, \citet{ma2011pair}, \citet{birk2012temperature}, \citet{furtenbacher2012marvel}, \citet{lodi2012line}, and \citet{tennyson2013iupac} }, using line broadening parameters appropriate for air. Since we are studying detectability of the band at relatively low spectral resolution $R\sim~140$, detailed line positions and other parameters are not of foremost importance.  We note that updated H$_2$O opacities are available \citep{Polyansky2018}, although we expect little change at these modest temperature conditions, so we have prioritized consistency with our previous work.

Water molecules exhibit three vibrational modes ($\nu_1$, $\nu_2$, and $\nu_3$). Those rovibrational transitions in which the quantum numbers change for two or more modes are called combination bands \citep{bernath2015spectra}. There are several combinations bands at spectral region 0.94$\mu$m, such as 2$\nu_1$+$\nu_3$, 1$\nu_1$+2$\nu_2$+1$\nu_3$. According to HITRAN \citep{gordon2017hitran2016}, the strengths of total individual lines in some bands such as 2$\nu_1$+$\nu_3$ are much larger than other bands and therefore they have the most impact on the opacity value. Other weak bands, however, were included in computing the water opacity in order to generate the water continuum accurately.

Following the albedo model setup of \citet{feng_robinson_fortney_lupu_marley_lewis_macintosh_line_2018}, we generated spectra of rocky exoplanets using Earth-like surface and atmosphere conditions as detailed in Table~\ref{table:params}.  The values chosen for these parameters produce realistic Earth spectra, as validated by \citet{feng_robinson_fortney_lupu_marley_lewis_macintosh_line_2018} against the NASA Astrobiology Institute's sophisticated 3D, line-by-line, multiple scattering Virtual Planetary Laboratory spectral Earth model tool  \citep{robinson2011earth}.   Four such spectra were generated, with atmospheric water content\footnote{Changes in H$_2$O content were compensated by changes in background N$_2$ gas content. See Section~\ref{sec:methods:retrieval} for more detail. All other parameters were left unchanged; see Section~\ref{sec:disc} for a discussion of possible ramifications of this choice.} of $0.01$, $0.1$, $1$, and $10$ times the Earth-like value described in Table~\ref{table:params}.  All models were generated with phase angle $\alpha=0$ -- full phase -- for this initial study.  Although true direct-imaging missions will not obtain full-phase observations, this assumption does not impact our results, as we do not compute integration times but instead work only in S/N space. Further, we anticipate performing a future followup investigation to expand this study and retrieve phase information.

\begin{figure}[!]
    \includegraphics[width=3.3in]{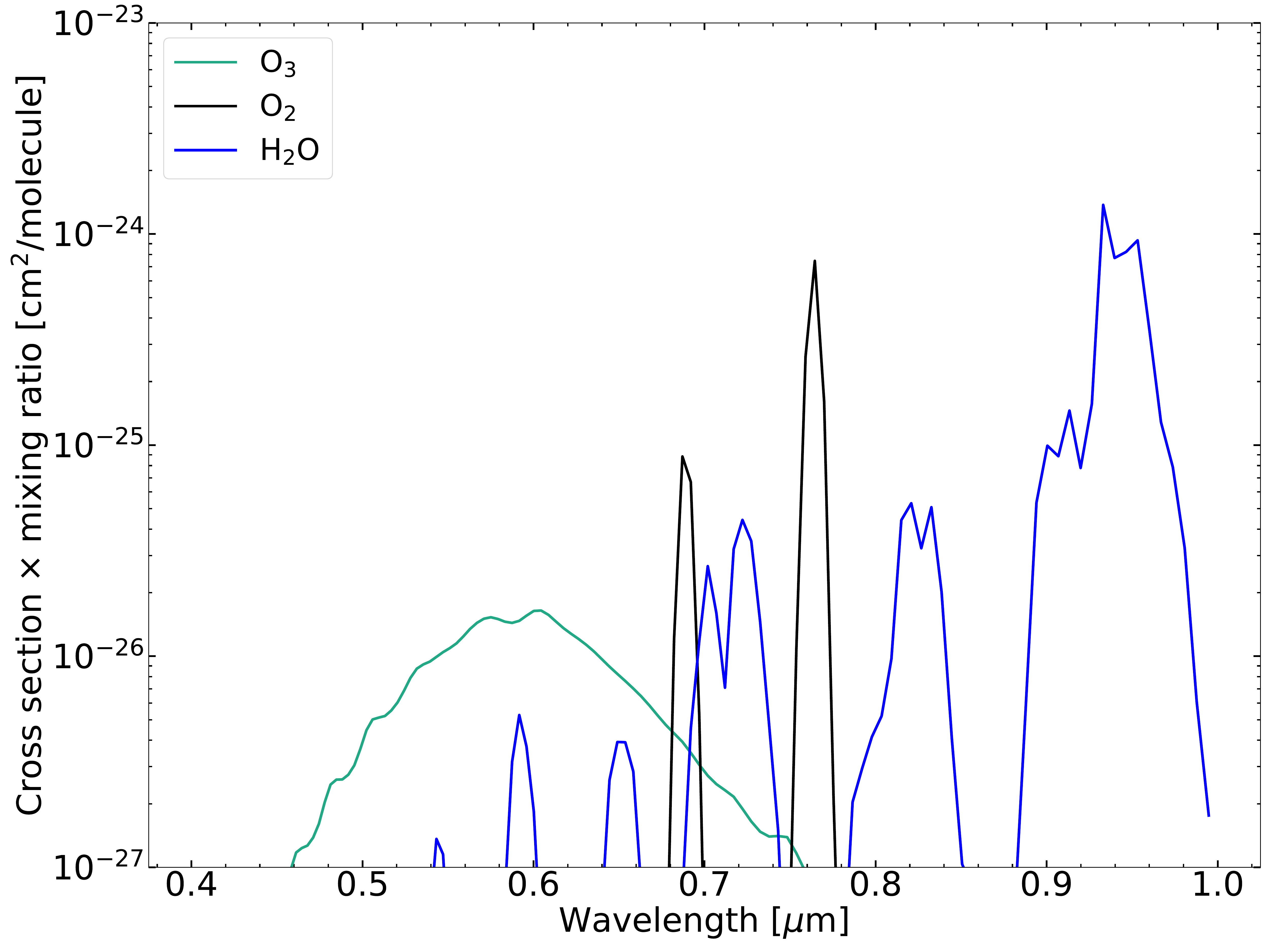}
    \caption {Weighted opacities represented logarithmically as a function of wavelength for O$_2$, O$_3$, and H$_2$O, the three major species discussed in this work. Each molecule has been weighted according to their relative abundance in our fiducial model. This opacity information manifests as absorption features in the contiuum albedo spectrum set by Rayleigh scattering and scattering from the grey clouds and surface. Features of interest are the wide, but shallow, O$_3$ absorption around 0.6\,$\mu$m, as well as the sharp O$_2$ features. The strong H$_2$O feature at 0.94\,$\mu$m is the primary target of this study. }
    \label{fig:opacities}
\end{figure}

Figure~\ref{fig:features} shows an example albedo spectrum at spectral resolution of $R=140$, shown as a planet-to-Sun flux ratio.  The most prominent features are due to Rayleigh scattering in the blue, a broad O$_3$ absorption, weak O$_2$ absorption, and water vapor features that grow in strength at redder wavelengths.  Figure~\ref{fig:opacities} shows the absorption cross-sections of these three important molecules, weighted by the mixing ratios of molecules in our standard ``Earth-like" setup.  While Rayleigh scattering imparts a slope in the blue, the optically thick water cloud is a gray scatterer throughout the rest of the optical.  The broad absorption due to O$_3$ gives a subtle dip in the spectrum   around 0.6\,$\mu$m, punctuated by narrower features due to O$_2$ and H$_2$O.

\begin{figure}[!]
    \includegraphics[width=3.3in]{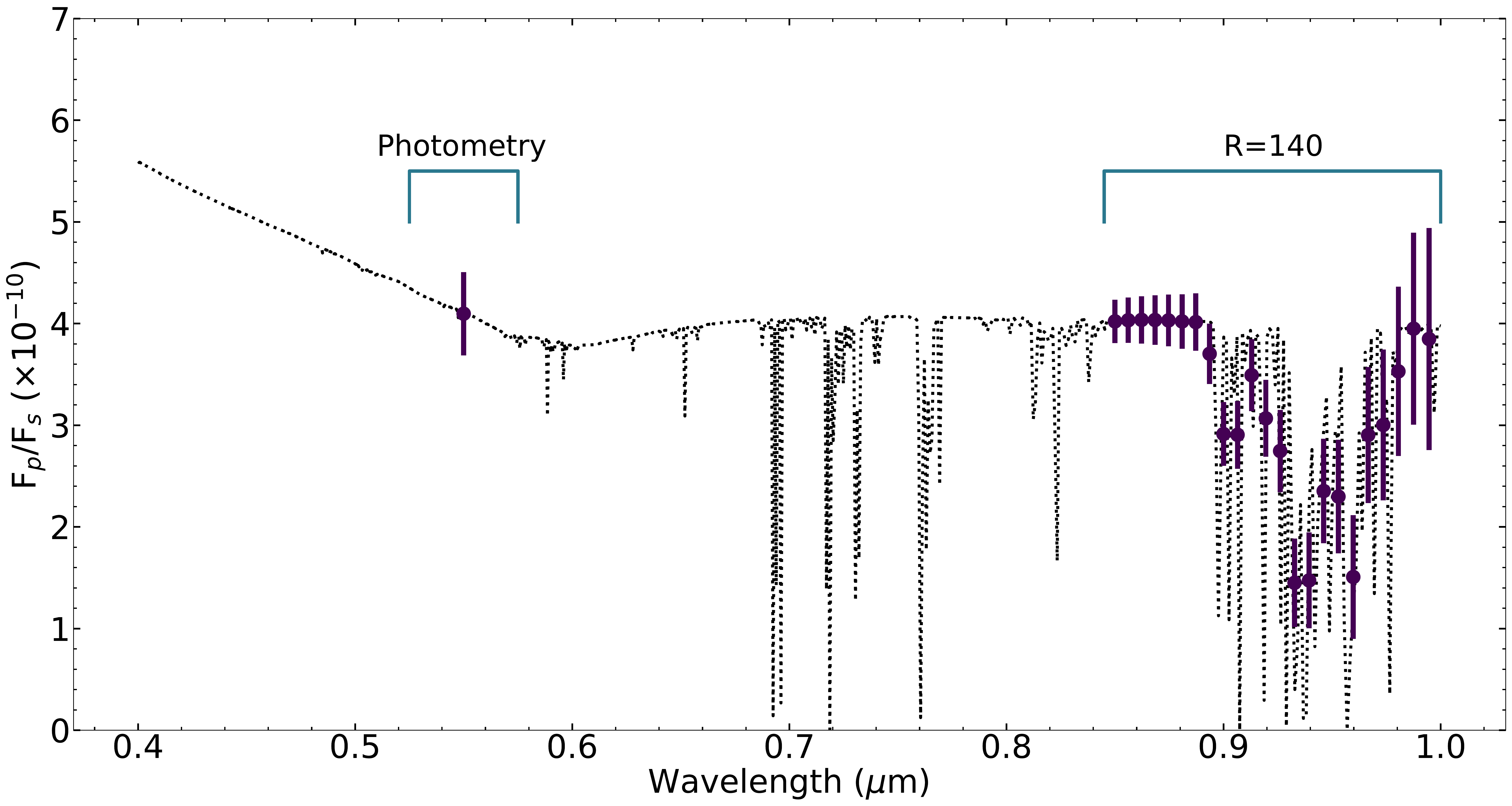}
    \caption{An example data set that was generated with an Earth-like water abundance, including the  high-resolution fiducial model} spectrum (black dotted line).  The region marked ``Photometry'' indicates the area covered by an potential optical photometric filter with a 10\% wavelength bandpass centered at 550 nm.  The region marked ``R=140'' indicates the area covered by an R$\,=140$ spectroscopic instrument with a 15\% bandpass, representing an immediate followup observation.  For this figure, we also used a signal-to-noise ratio of 15 in this region, calculated at $0.88 \,\mu$m.
    \label{fig:datapoints}
\end{figure}

\subsection{Observation Simulation}

With high-resolution albedo spectra in hand, we next simulated observations of these objects with a coronagraph-equipped telescope.  The basic idea was to generate data that may be akin to some ``first'' observations, including a broadband optical photometric point (for planet discovery), followed by a reconnaissance spectrum across the $0.94\,\mu$m water band.  This was achieved by reducing the resolution of the simulated spectrum to produce data points with spectral resolution R$\,=140$ in a 15\% bandpass centered on the strong H$_2$O absorption feature at $0.94\,\mu$m (see Figure~\ref{fig:features}), covering the region from $0.85-1.00\,\mu$m.  We chose this range so as to include the water absorption feature along with ``continuum'' reflection off the clouds, just blue-ward of the water band.  What we term continuum reflection is the relatively gray reflection from the optically thick water clouds, from $\sim$~0.6 to 1.0$ \,\mu$m, punctuated only by O$_2$ and H$_2$O absorption.

We combined this medium-resolution data with an integrated $0.525\,\mu$m - $0.575\,\mu$m photometric point (termed elsewhere in this paper as a 0.55$\,\mu$m or ``green'' data point).  We employed the noise model of \citet{robinson2016characterizing} to simulate signal-to-noise ratios (SNR) achieved by the instrument, where the signal is defined as the reflected flux ratio $F_{\mathrm p}/F_{\mathrm s}$.  For each chosen value of SNR, we selected a ``continuum" reference data point at $\lambda_0 = 0.88\,\mu$m, just outside of the water band, and set the uncertainty of that data point to be $\Delta F_{\mathrm p}/F_{\mathrm s}$ = $F_{\mathrm p}/F_{\mathrm s}$/SNR.  Equation 6 of \citet{robinson2016characterizing} relates exposure time to background photon count rate, planet photon count rate, and signal-to-noise ratio.  Although background count rate, planet photon count rate, and signal-to-noise ratio may not be constant across all wavelengths, exposure time must be a constant within a single bandpass; therefore, we may equate the value at the reference wavelength $\lambda_0$ with that at another wavelength $\lambda$ by

\begin{equation}
    \frac{c_p(\lambda_0)+2c_b(\lambda_0)}{c_p(\lambda_0)^2} SNR(\lambda_0)^2 = \frac{c_p(\lambda)+2c_b(\lambda)}{c_p(\lambda)^2} SNR(\lambda)^2
\end{equation}

We can therefore solve for the wavelength-dependent signal-to-noise ratio $SNR(\lambda)$, allowing us to extrapolate the uncertainties achieved at all data points, given a set signal-to-noise ratio at the reference data point, a model albedo spectrum, and a model of background photon counts.  We use \citet{robinson2016characterizing} to inform our background photon counts, the Cahoy et al.~model described above to produce albedo spectra, and values of 5, 10, and 15 for our reference data point signal-to-noise ratios.  See Figure~\ref{fig:datapoints} for an example data set.

Following \citet{feng_robinson_fortney_lupu_marley_lewis_macintosh_line_2018} we add these appropriate error bars to the reduced-resolution fiducial model spectrum, but do not randomize the data points.  As discussed in \citet{feng_robinson_fortney_lupu_marley_lewis_macintosh_line_2018} this is a choice of convenience, but with a purpose.  The retrieval on a single noise instance could easily bias our retrieval results.  The retrieval on a large number of noise instances would be most proper, but is computationally extremely expensive.  From tests \citet{feng_robinson_fortney_lupu_marley_lewis_macintosh_line_2018} demonstrated that posteriors on atmospheric quantities of interest, comparing non-randomized data and that achieve from 10 different noise instances, led to good qualitative agreement.  While acknowledging that our treatment here is likely modestly optimistic compared to a more detailed treatment, our work certainly show important trends that set a basis for more comprehensive followup work.

Our adopted spectral resolution (R$=140$) is consistent with both the current HabEx and LUVOIR designs at wavelengths around the 0.94~$\mu$m water vapor spectral feature.  Proposed coronagraphs for both the HabEx and LUVOIR concepts would achieve corongraph bandwidths of 10--20\%, which is consistent with our adopted bandwidth (15\%).  We note that the primary HabEx design also includes a starshade capable of performing high-contrast imaging and spectroscopy across the full 0.45--1.0~$\mu$m range in a single pointing, which would supersede the bandpass adopted here.  Finally, our study explores retrievals at different characteristic SNRs so as to avoid tying our results to a specific telescope design.  Nevertheless, our SNRs can be converted to requisite integration times for the HabEx and LUVOIR concepts using available instrument models\footnote{https://habex.ipac.caltech.edu/}\footnote{https://asd.gsfc.nasa.gov/luvoir/tools/} or through the \citet{robinson2016characterizing} noise model.\textbf{}


\subsection{Retrieval} \label{sec:methods:retrieval}

Simulations were produced by pairing our albedo and noise models, which we then treated as observational data, and a Bayesian retrieval was performed using the PyMultiNest software \citep{feroz_hobson_bridges_2009, buchner2014x}, following \citet{feng_robinson_fortney_lupu_marley_lewis_macintosh_line_2018}.  Computational Bayesian retrieval techniques involve the comparison of model outputs to the observed data, using a variety of algorithms (such as Markov-chain Monte Carlo, or the Multi-Nested methods used here) to explore the parameter space in an efficient manner.  By quantifying the ``likelihood'' of each set of parameters producing the observed data, we seek to understand the distribution of possible values of those parameters. By comparing these posterior distributions to the known input values for each parameter, we can determine the information content of a data set at a given signal-to-noise ratio.  We have previously used these techniques in both gas giant and terrestrial planet reflection spectra  \citep{lupu_marley_lewis_line_traub_zahnle_2016,feng_robinson_fortney_lupu_marley_lewis_macintosh_line_2018}.

The \citet{cahoy_marley_fortney_2010} albedo model used in this project accepts as input the 13 parameters detailed in Table~\ref{table:params}: mixing ratios for five molecules: ozone (O$_3$), oxygen (O$_2$), methane (CH$_4$), carbon dioxide (CO$_2$), and water (H$_2$O); surface properties of atmospheric pressure ($P_0$), gravitational acceleration ($g_{\mathrm{pl}}$), and surface reflectivity ($A_{\mathrm s}$); the planetary radius ($R_{\mathrm{pl}}$); and four cloud properties: cloud top pressure ($p_{\mathrm t}$), cloud pressure thickness ($\delta p$), cloud optical depth ($\tau$), and cloud coverage fraction ($f_{\mathrm{cld}}$).  By repeatedly sampling values of each retrieved parameter and comparing the albedo spectrum output to calculate a numerical likelihood, the Bayesian retrieval tool builds up a posterior probability distribution for all ``free'' retrieved parameters.

\begin{figure}[!]
    \centering
    \includegraphics[width=3.0in]{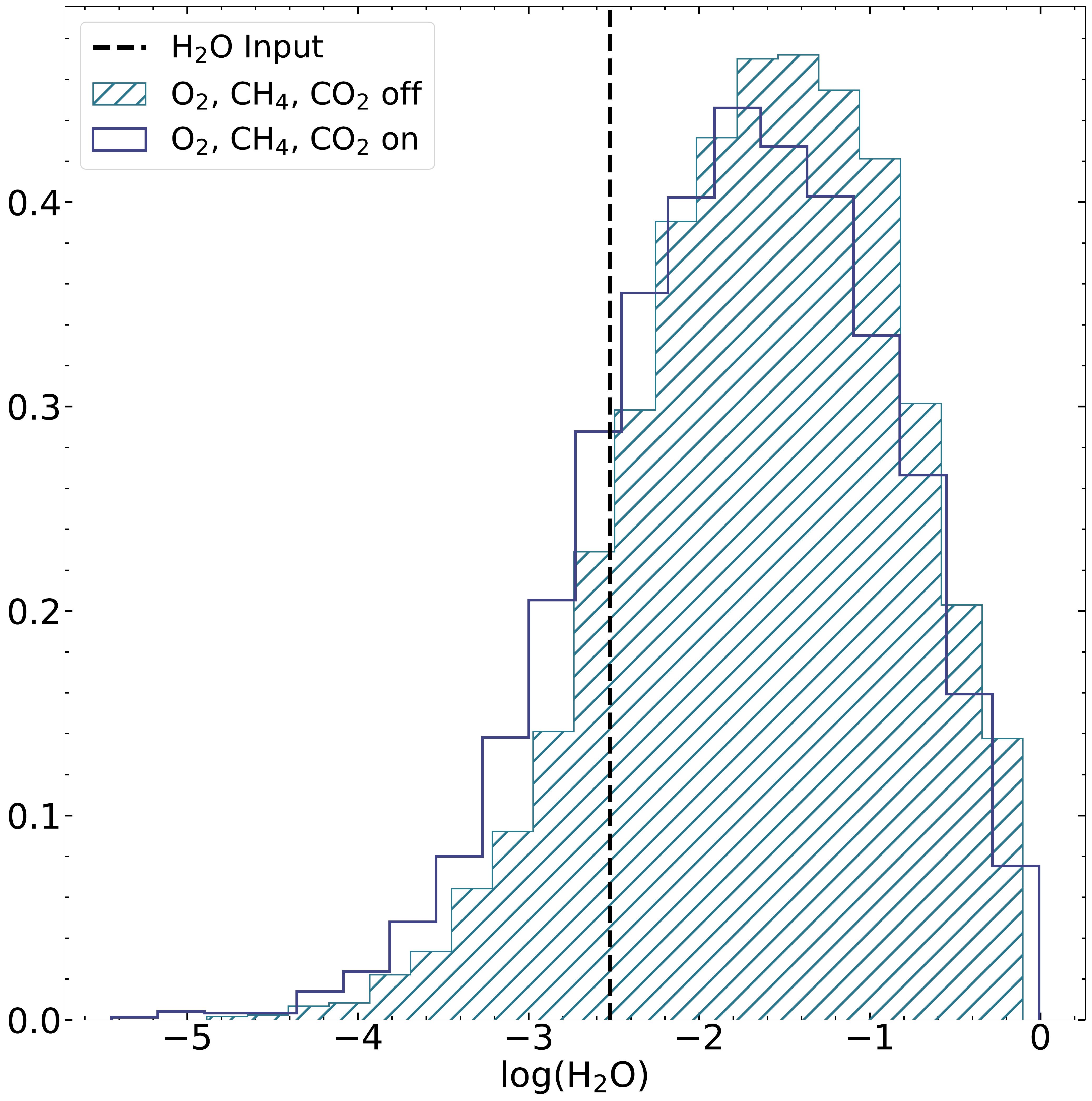}
    \caption {An illustration of the impact of including molecular oxygen O$_2$, methane CH$_4$, and carbon dioxide CO$_2$ as free parameters in retrievals.  These two posterior distributions of H$_2$O were retrieved from a data set based on a forward model with $1\times$ Earth-like water mixing ratio and a signal-to-noise ratio of 15.  Because of the very small difference in the posteriors, we chose to fix these parameters in order to reduce retrieval times.}
    \label{fig:molecules}
\end{figure}

During this process, we chose not to retrieve for the abundances of molecular oxygen O$_2$, methane CH$_4$, and carbon dioxide CO$_2$, instead fixing them at the values presented in Table \ref{table:params}.  As shown in Figure~\ref{fig:features}, there are no major features for any of these three molecules in the region of interest.  Extensive testing revealed that the retrieval posteriors for the molecules in question were uninformative when they were included in free parameters.  H$_2$O posteriors retrieved with and without fixing the abundances of these molecules were virtually indistinguishable, as shown in Figure \ref{fig:molecules}, where we compare retrievals performed with and without these gasses. This test was conduced by performing two retrievals on the same data set: one in which these molecules were allowed to be free parameters, to be retrieved by the nested sampler; and a second, in which we held them fixed while retrieving for other parameters \emph{given} the values of the mixing ratio for these two molecules.  Since including these gasses leads to increased computational time, we elected to leave them fixed at truth values during the retrieval.

Beyond that depicted in Figure~\ref{fig:molecules}, a total of 19 retrievals were performed:  A ``primary'' set of twelve retrievals, and an ``auxiliary'' set of seven. The primary set explored planets with H$_2$O mixing ratios at 0.01$\times$, 0.1$\times$, 1$\times$, and 10$\times$ the current Earth levels, with signal-to-noise ratios of 5, 10, and 15, retrieving on the parameters as described above. When adjusting the H$_2$O mixing ratios, we held the mixing ratios of the other spectrally active atmospheric constituents fixed by increasing (decreasing) the background N$_2$ gas ratio to compensate for the decreased (increased) H$_2$O presence.

Three of the auxiliary retrievals were performed on data sets were derived from a $1\times$ Earth-like model, using SNR$\,=5, 10,$ and 15, but \emph{without} the $0.525-0.575\,\mu$m optical photometric point.  These were analyzed with our retrieval framework in order to explore the value of this green data point indirectly, by examining the information contained only in the red spectroscopic observation.  While in general the $0.525-0.575\,\mu$m optical photometric point is expected to be available, understanding the effect of its absence allows us to consider cases where there is concern about the validity of the optical point data for any reason; for instance, if there was a significant time delay between the photometric detection and followup spectroscopy, or that the phase angles of the two measurements may differ.

The final set of four auxiliary retrievals were performed on data sets that used 0.10 and $0.15\,\mu$m wide photometric \emph{filters} over the $0.94\,\mu$m absorption band (as well as the optical photometric point) for model planets with $1\times$ and $10\times$ Earth-like water mixing ratios.  The rationale was to understand if \emph{photometry} could give any constraint on water vapor, instead of more time-consuming spectroscopy.

\section{Results} \label{sec:results}

Although we retrieved on many atmospheric properties in our runs, the H$_2$O mixing ratio was of primary interest.  Therefore, the results of H$_2$O mixing ratio retrieval will be shown in some detail.  Results for other parameters where meaningful constraints could be placed will also be discussed.

Following \citet{feng_robinson_fortney_lupu_marley_lewis_macintosh_line_2018}, we define the following 4 terms for use when discussing our results.

\begin{enumerate}
    \item A \emph{non-detection} describes a retrieved posterior that is flat or nearly flat across the entire prior range.
    \item A \emph{weak detection} describes a retrieved posterior that shows a peak, but has significant ``tails" going to one or both ends of the prior.  This includes retrievals that produce upper or lower limits on the parameter.
    \item A \emph{detection} describes a posterior that shows a localized peak without significant ``tails", but that have a $1\sigma$ range greater than one order of magnitude.
    \item A \emph{constraint} describes a peaked posterior distribution, similar to a detection, with a $1\sigma$ range of less than one order of magnitude.
\end{enumerate}

\begin{figure}[!]
    \includegraphics[width=3.0in]{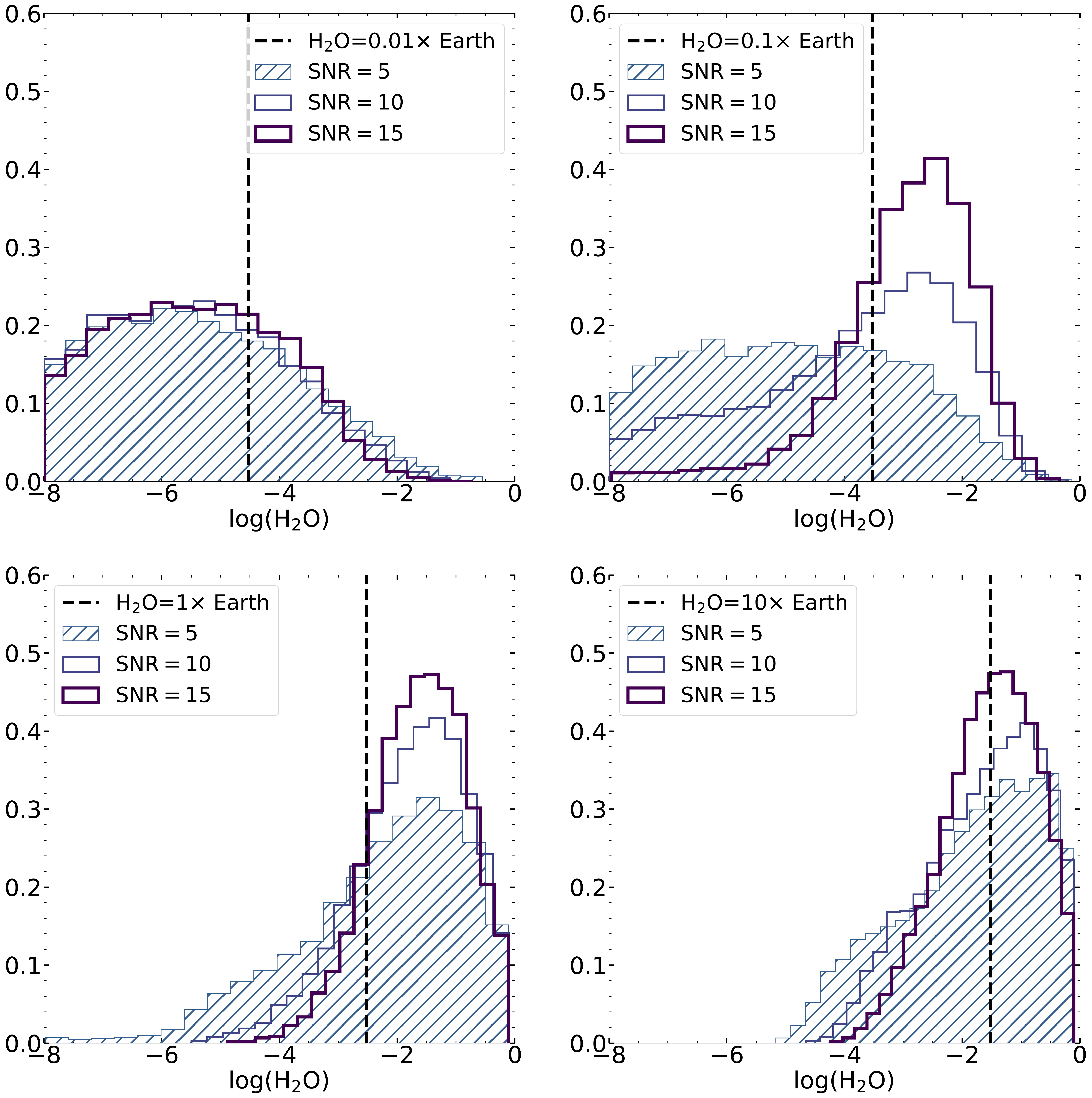}
    \caption{H$_2$O posteriors for data sets with H$_2$O $=0.01\times$ (top left), $0.1\times$ (top right), $1\times,$ (bottom left) and $10\times$ (bottom right) Earth-like values. Each data set combines a 0.55\,$\mu$m photometric point (with fixed signal-to-noise ratio of 10) and R = 140 spectroscopy centered on the 0.94\,$\mu$m water band. For each water mixing ratio, we vary the signal-to-noise ratio of the spectrum: SNR = 5 (teal, cross-hatched), 10 (thin purple line) and 15 (thick magenta line) are shown, along with the location of the truth value (black dashed line). Water is weakly detected even for the lowest SNR and lowest mixing ratio. To claim detection, there needs to be at least 0.1$\times$ Earth water in the atmosphere and corresponding SNR = 15 data; for water content $1\times$ Earth value, the data need SNR = 10 for detection. SNR = 15 offers constraint for H$_2$O = $1\times$\ and $10\times$\ Earth values only. }
    \label{fig:h2ostack}
\end{figure}

Our standard setup included the $0.55\,\mu$m optical point with SNR$\,=10$, and an R$\,=140$ spectrum across the $0.94\,\mu$m water band.  In retrievals with H$_2$O $=0.01\times$ Earth-like values, we found that all SNR tested resulted in weak detections of H$_2$O mixing ratios. In retrievals with H$_2$O $=0.1\times$ Earth-like values, we also found that low SNR resulted in weak detections of H$_2$O. In particular, the SNR$\,=5$ retrieval in this regime resulted in a detection weak enough to be considered a non-detection. However, with SNR$\,=15$ we achieved a detection.  In retrievals with H$_2$O $=1\times$ Earth-like values, we achieved a weak detection with SNR$\,=5$, a detection with SNR$\,=10$, and a constraint with SNR$\,=15$.  Similarly, in retrievals with H$_2$O $=10\times$, we achieved detections with SNR$\,=5$ and $10$, and a constraint with SNR$\,=15$.  These results are summarized in Figure~\ref{fig:h2ostack}.

In our retrievals we were only able to constrain -- or, indeed, achieve a detection -- on a few atmospheric parameters, including the H$_2$O mixing ratio, $P_0$ (surface pressure), $p_{\mathrm t}$ (cloud-top pressure), and $\delta$p (cloud thickness in pressure).  Surface pressure moved from a weak detection to a detection at H$_2$O $=10\times$ Earth-like values and SNR$\,=10$, and at all water abundances at SNR$\,=15$.  The cloud properties p$_t$ and $\delta$p moved from weak detection to detection at H$_2$O $=0.1\times$ Earth-like values and greater with SNR$\,=10$, and at all water abundances at SNR$\,=15$.  The cloud properties $\tau$ and f$_{cld}$ returned non-detections in all retrievals.  $A_s$ was undetected at H$_2$O $= 0.1\times$ Earth-like values and lower with SNR$\,=5$ and $10$, and weakly detected in other retrievals.  R$_{pl}$ and g$_{pl}$ are weakly detected in all retrievals.

\begin{figure}[!]
    \includegraphics[width=3.3in]{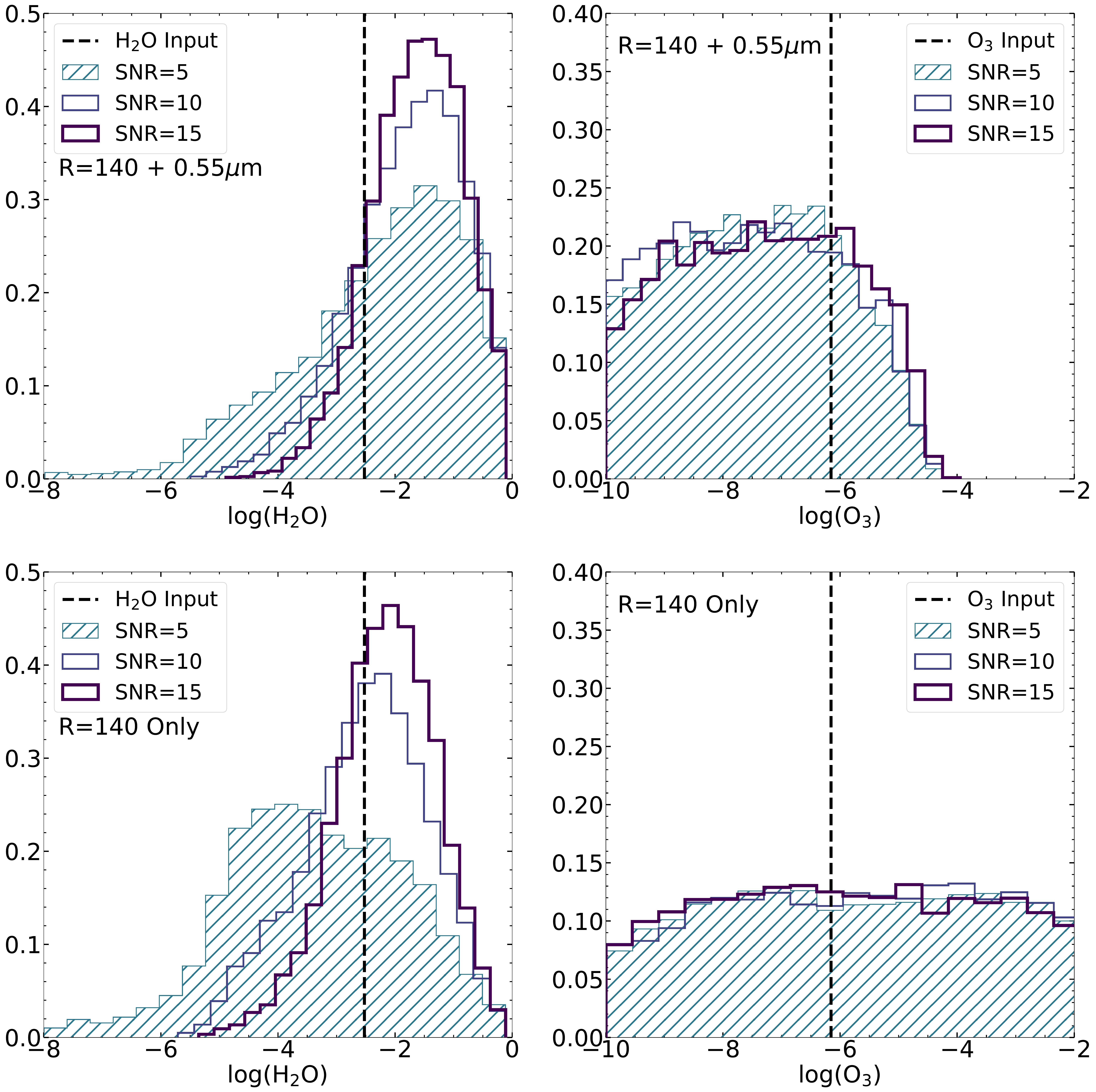}
    \caption{Three retrievals were performed on a data set which did not include an optical 0.55\,$\mu$m photometric data point.  These were done with H$_2$O $=1\times$ Earth-like values and SNR $=5, 10,$ and $15$.  Top row: we show the results of those retrievals. Bottom row: we plot the results of retrievals on data with the same H$_2$O and SNR values that does include the optical photometric data point. While the posteriors of water do not change much, this shows that the inclusion of the photometric point allows us to go from a non-detection to placing an upper limit on O$_3$.}
    \label{fig:optival}
\end{figure}

Of additional note is that we did achieve a weak detection of O$_3$ with an upper bound in even our poorest SNR retrievals.  Evidently, the green photometric discovery point has some utility in placing an upper limit on ozone.  In the interest of determining the value of a same-phase optical photometric point in the retrieval of H$_2$O, we also performed retrievals with SNR$\,=5,10,$ and $15$ and H$_2$O $=1\times$ Earth-like values with this point not included, shown in Figure~\ref{fig:optival}.  Although this modestly degraded the precision and accuracy of the retrieval for all retrieved features, particularly for the lowest SNR case, it did not substantially change the shape of the posterior for H$_2$O in the SNR$\,=10$ and 15 cases.  In addition, we lost the upper limit on O$_3$ and were left with a non-detection of ozone in all retrievals.

\begin{figure}[!]
    \centering
    \includegraphics[width=3.3in]{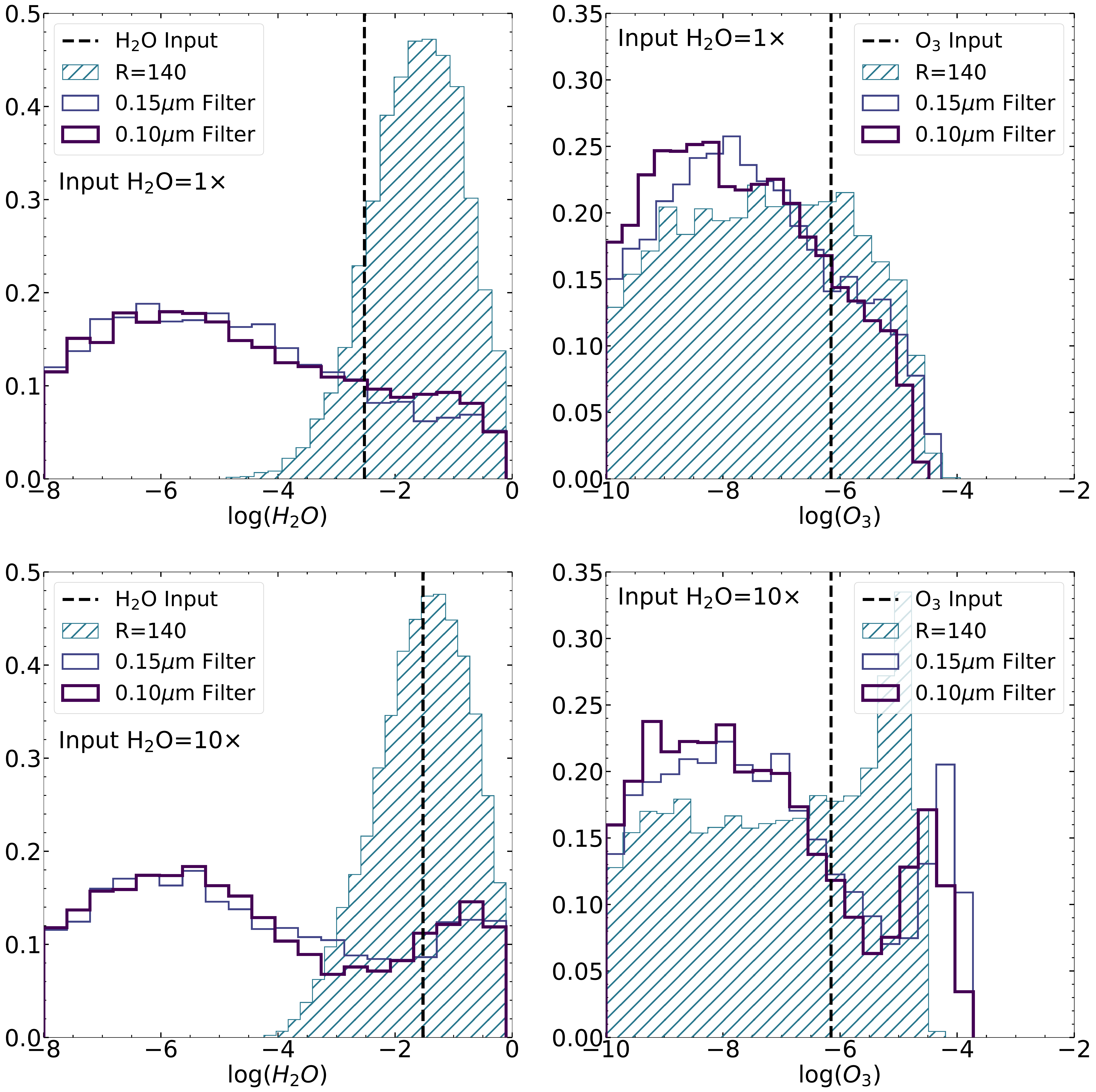} 
    \caption{Retrieved H$_2$O posterior distributions for photometric data compared with an R$\,=140$ spectrum.  Retrieval results for H$_2$O (left) and O$_3$ (right) with truth H$_2$O values of $1\times$ (top) and $10\times$ (bottom) Earth-like values. We compare the R$\,=140$ scenario (teal cross-hatch) used in the rest of this paper with two scenarios employing photometric filters across the $0.94\,\mu$m H$_2$O absorption feature.  One scenario used a $0.15\,\mu$m filter (thin purple line) from $0.85-1.00\,\mu$m.  The second scenario used a narrower, $0.10\,\mu$m filter (thick magenta line) from $0.90-1.00\,\mu$m. All posteriors shown in this figure were retrieved from SNR=15 data. The switch from spectroscopy to photometry of the water band means that we would go from constraint of H$_2$O to non-detection even at SNR=15. The switch would not impact O$_3$ inference much, with all cases retrieving upper limits for the molecule.}
    \label{fig:specVfilter}
\end{figure}

As discussed above, an option considered for H$_2$O detection, which would require the least integration time, was the use of a photometric filter centered on the H$_2$O feature at $0.94 \,\mu$m.  Therefore, we performed four retrievals under this assumption.  We considered Earth-like and $10 \times$ Earth-like H$_2$O concentrations, which we judged to be the most favorable cases given our retrievals with R$\,=140$ spectroscopy. We tried two different filter widths of $0.85-1.00\,\mu$m and $0.90-1.00 \,\mu$m and combined each water band photometric point with the optical point at 0.55\,$\mu$m to form data sets with SNR = 15.  However, we retrieved non-detections on H$_2$O in all situations, as shown in Figure~\ref{fig:specVfilter}, which shows posteriors for water and ozone.  This suggests that even well-place photometric points will be of little aid in classifying rocky planets as a tool to decide on future detailed characterization.  This echoes the finding of \citet{batalha2018color} in their exhaustive study of giant planet albedos.

\section{Discussion and Conclusions} \label{sec:disc}

A ``follow the water'' strategy for terrestrial exoplanet atmospheric characterization may be a useful one for determining which worlds may be most interesting for detailed follow-up observations.  Through some initial retrieval explorations that used simulated observations from a large space-based telescope, we have been able to start shedding some light on how this might be best accomplished.  First, while filter photometry allows for shorter integration times to achieve a given SNR, and will likely be how the first planet detections are made, it will be of limited aid in characterizing atmospheric water abundances.

In addition, the diagnostic power of spectroscopy was so high that retrievals for the water mixing ratio, from spectroscopy, were not particularly aided by the additional of a ``discovery phase'' green photometric point.  While the presence or absence of this data point did have an effect on the precision of our retrieval results, we found that with a signal-to-noise ratio of 10 or higher the precision of retrieved water abundance was sufficiently comparable to results that included the data point that we do not expect it would have a substantial impact on our overall conclusions.  The 2$\sigma$ lower bounds without the green photometric point are $10^{-0.5}\times, 10^{-1.1}\times,$ and $10^{-2.2}\times$ Earth-like values for SNR$\,=5, 10,$ and 15, respectively.  These values are essentially similar to those found for retrievals which included the green photometric point.  It is worth noting, however, that this data point did allow for upper limits to be placed on ozone abundances. 



\begin{figure}[!]
    \centering
    \includegraphics[width=3.3in]{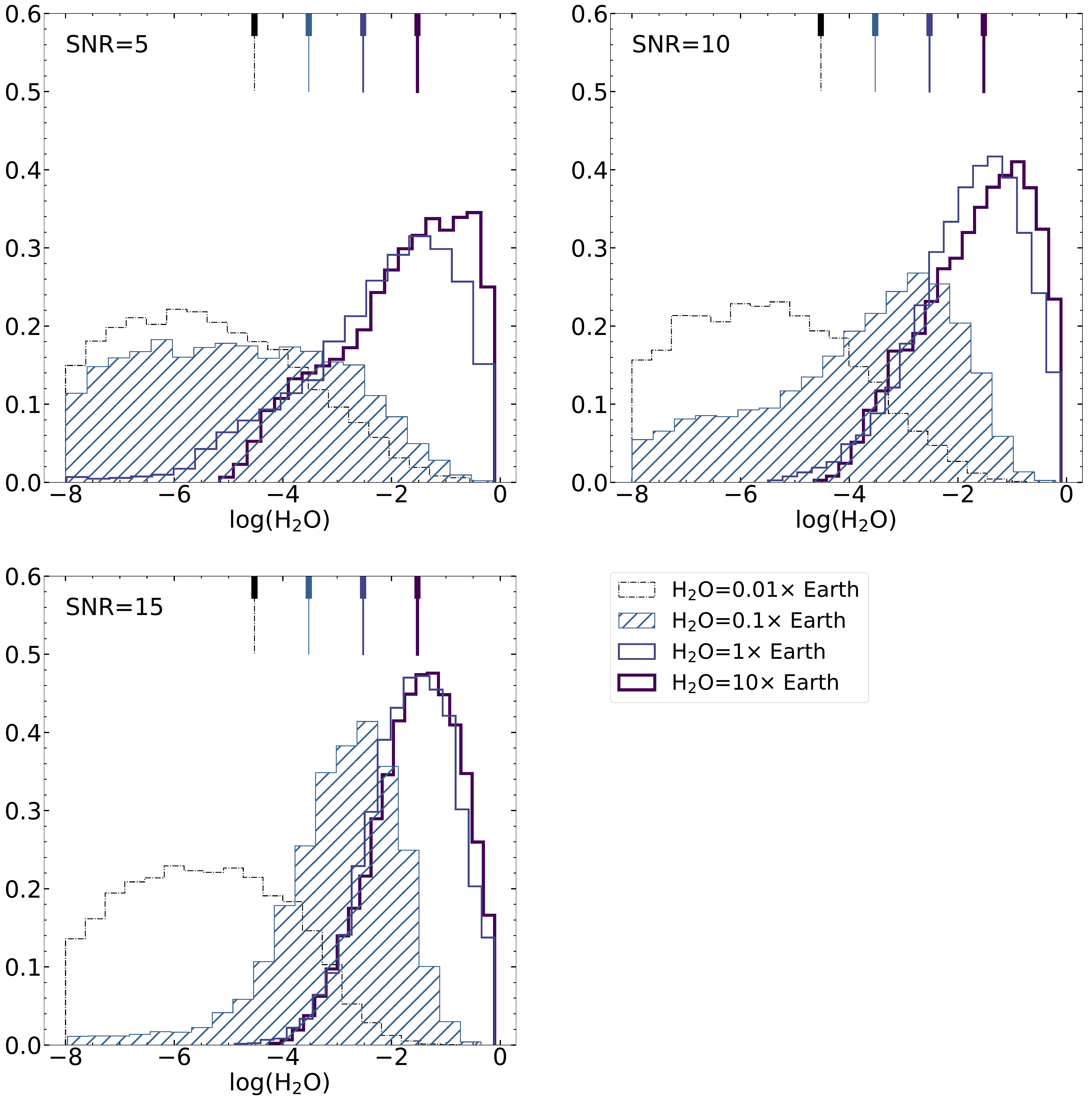} 
    \caption{These H$_2$O posteriors are identical to those shown in Figure~\ref{fig:h2ostack}, but sorted by signal-to-noise ratio in order to show trends and the ability to distinguish planets based on atmospheric water content.  The vertical marks along the top of the image indicate the four truth values used in this study.  We see that a retrieval on SNR=5 data results in substantial differences in the posterior between high ($\geq 1\times$ Earth-like) and low ($\leq 0.1\times$ Earth-like) water mixing ratio objects, while at SNR=10 and 15, additional differences begin to emerge for 0.1$\times$ objects. For the driest case, increasing SNR does not improve detection beyond an upper limit. For planets with at least $0.1\times$\ Earth water mixing ratio, we can benefit from better SNR data as detection is possible. If a terrestrial planet's atmosphere is sufficiently abundant with water ($\geq 1\times$ Earth-like), we can detect its presence with SNR = 5.}
    \label{fig:snrstack}
\end{figure}

Using an R$\,=140$ spectrum from $0.85-1.00\,\mu$m, in conjunction with the green optical point, constraints on H$_2$O abundance were only possible in high signal-to-noise ratio cases, and then only with a significant water presence.  However, if constraints on abundances are not necessary for early atmospheric characterization, it may be helpful to consider the requirements to distinguish the water-bearing worlds from the dry worlds, as the presence of water vapor in any significant quantity may indicate a world of interest to astrobiological studies.  In Figure~\ref{fig:snrstack} we have re-plotted the data from Figure~\ref{fig:h2ostack} to show the appearance of the H$_2$O posterior with different truth H$_2$O values, while keeping the signal-to-noise ratio constant.

Although our retrievals were not able to place strong constraints on H$_2$O mixing ratios, we did find that a low SNR$\,=5$ retrieval may have some utility to distinguish worlds with H$_2$O $\geq 1\times$ Earth-like levels from those with H$_2$O $\leq 0.1\times$ Earth-like levels.  These SNR$\,=5$ retrievals yielded uninformative $2\sigma$ lower water mixing ratio limits\footnote{ $10^{-3.5}\times$ and $10^{-4.3}\times$ Earth-like levels, respectively. Both of these values approach the lower limit of the sampled range. } for $0.01\times$ and $0.1\times$ Earth-like water mixing ratios, while retrievals on Earth-like and greater water mixing ratio models returned $2\sigma$ lower-bound values in excess of $0.3\times$.  Additionally, with the higher SNR$\,=10$, the $0.1\times$ worlds begin to become distinguishable from $0.01\times$ worlds, as the $2\sigma$ water mixing ratio lower-bound for $0.1\times$ worlds rises to $10^{-2.5}\times$ Earth-like values, while that of $0.01\times$ worlds remains unchanged.  These distinctions are possible because the retrieved posterior distributions show a strong sensitivity to H$_2$O mixing ratio.  $2\sigma$ upper limits were largely uninformative in all retrievals, with useful values only becoming apparent in high SNR$\,=15$ retrievals on very low H$_2$O models.  We caution that the $0.01\times$ models were not distinguishable from essentially dry worlds even at SNR$\,=15$.

As outlined above, our work builds on that of  \citet{feng_robinson_fortney_lupu_marley_lewis_macintosh_line_2018}, who applied Bayesian retrieval techniques to simulated observations of model true-Earth analogs, but with different observational assumptions. Thus a comparison between these two studies may be prudent.  When  \citet{feng_robinson_fortney_lupu_marley_lewis_macintosh_line_2018} studied R=140 spectroscopy of the full optical spectrum from $0.4-1.0\,\mu$m, much stronger retrieved detections were produced on atmospheric parameters that we have no handle on in this work.  This is unsurprising, given our narrow wavelength range of interest.  A similar story emerges when comparing H$_2$O, specifically; while neither the present work nor the \citet{feng_robinson_fortney_lupu_marley_lewis_macintosh_line_2018} R=140 retrievals were able to make strong statements about H$_2$O mixing ratio with a signal-to-noise ratio of 5, at high signal-to-noise ratios this work fared substantially poorer in both precision and accuracy of retrieved values.  We also note that while we used a SNR reference $\lambda_0=0.88\,\mu$m, which is appropriate for our study, \citet{feng_robinson_fortney_lupu_marley_lewis_macintosh_line_2018} used instead a value of $\lambda_0=0.55\,\mu$m, which was appropriate for the wide optical wavelength range they studied.


\begin{figure}[!]
    \centering
    \includegraphics[width=3.3in]{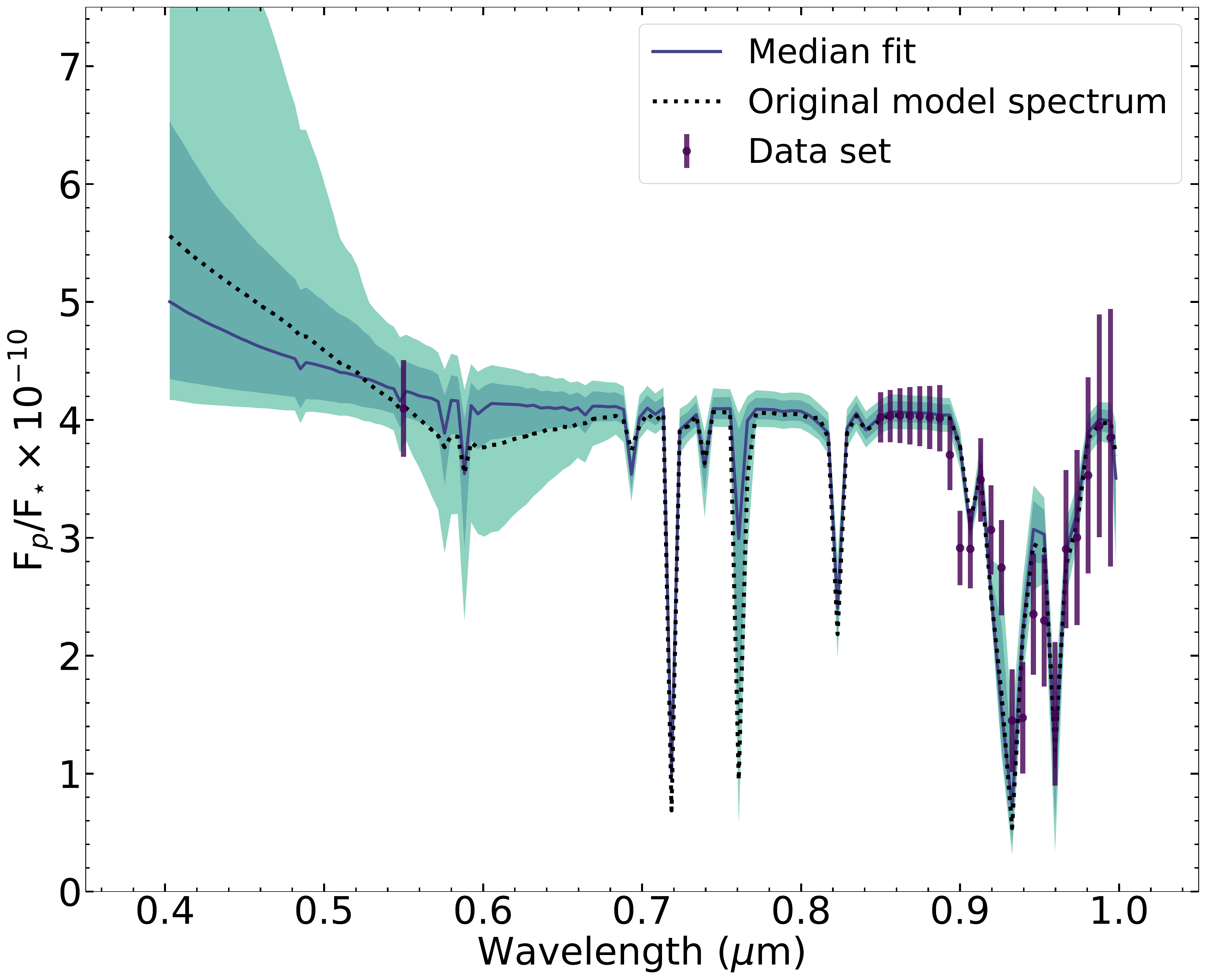} 
    \caption{In this image, the median (blue), 1$\sigma$ (dark green), and 2$\sigma$ (light green) parameter sets have been used to generate albedo spectra, and have been overplotted with the truth albedo spectra (black dotted line) and the input data set (purple). The regions with no data show a large spread in possible flux ratio values. Additional spectroscopic coverage can ensure better constraint of the atmospheric state and planetary properties such as surface albedo and surface pressure.}
    \label{fig:yarr}
\end{figure}

Another way to appraise our results is to consider how well the albedo spectrum of the planet is constrained at wavelengths outside of where data are obtained, operating within our framework of assuming a rocky exoplanet and using parameters based on our retrieval results.  This is shown in Figure~\ref{fig:yarr}.  As might be expected, the retrieved atmospheric parameters yield optical spectra that tightly correspond to the fiducial model spectrum in each of the water features, but we see considerable deviation outside of these regions.  In particular, with no data across the 0.77 $\,\mu$m O$_2$ A-band, we visually see little constraint on the feature depth, compared to the excellent fit with with the weaker water features to the blue and red of this O$_2$ band.  A future investigation might look at data near the O$_2$ A-band, perhaps including the H$_2$O $\alpha$-band absorption feature at 0.72 $\,\mu$m within a $\sim$~10-15\% bandpass.  Another path this suggests for future studies is comparing results with other, non-Earth-like models -- a retrieval where one is truly blind to atmosphere type may suggest a path to differentiate water-bearing Earth-like worlds from water-bearing small sub-Neptunes.

Taking all this together, a picture begins to emerge as to the value of the  combined R=140 spectroscopy and 0.525-$0.575\,\mu$m photometry data collection setup used in this study.  While clearly insufficient as a means for detailed characterization, using spectroscopy on the $0.94\,\mu$m feature appears to be a useful method for quickly distinguishing between wet and dry rocky exoplanets. This distinction can then be used to guide broader, more time-intensive followup studies in a search for life-bearing exoplanets.  We note, however, that a photometric band centered on this same $0.94\,\mu$m water feature provided little utility, as even at a high SNR water vapor was not detected.  In addition, we found that making use of the $0.55\,\mu$m optical data point can allow one to place some constraints on O$_3$, and at high-SNR using both can allow for determination of some useful information about the planet's surface pressure and some cloud properties.

Additional studies may be prudent for a better understanding of the limitations of this technique.  Here we only examined a single spectral resolution, of R$\,=140$.  Lower resolution could be explored across this relatively wide bandpass.  Furthermore, the H$_2$O mixing ratios studied here were each spaced by an order of magnitude, while other properties were left at Earth-like levels.  Studies with additional granularity in H$_2$O mixing ratio may provide some benefits, as would consideration of the other physical properties of the planet.

A more physically motivated ``Earth'' model could include additional physical effects.  It may reasonably be expected that altering the water mixing ratio will impact the properties of water clouds in the planetary atmosphere.  As a world with less water will, perforce, have fewer water clouds, an observation of such a world would see deeper into the atmosphere. This would, to some extent, strengthen the water vapor absorption feature, thus we expect to see some slight degeneracy between f$_{cloud}$ and log(H$_2$O).  The height in the atmosphere where water clouds reside would likely change as well, although a change in water mixing ratio would alter the greenhouse effect and hence the temperature structure of the atmosphere, including condensation levels.  While these effects may alter the particulars of the model results, it is not expected that the degeneracy between f$_{cloud}$ and log(H$_2$O) would be strong; further, given our retrievals' relative insensitivity to cloud features in general, and in particular to f$_{cloud}$ such changes are unlikely to impact the broader conclusions drawn by this study.

Finally, changes to water mixing ratios may impact other parameter values in a way which is not represented here. In particular, as explored in \citet{wordsworth2013water}, changes to water mixing ratios may impact carbon dioxide mixing ratios for Earth-like planets.  This was not modelled here, although as discussed, we did not find that our results were dependent on CO$_2$ mixing ratios given our focus on the optical bandpass.  Clearly, temperate rocky planets can present a wide range of atmospheric states, and much work lies ahead in assessing how to characterize these potentially habitable worlds.

\acknowledgments

The authors would like to acknowledge Xi Zhang for interesting discussions on atmospheres with high water mixing ratios and Bruce Macintosh for insights on the observational context.  We thanks the referee and Ehsan Gharib for helpful discussions on the clarity of the work and related opacity issues.  TDR, JJF, and MSM gratefully acknowledge support from an award through the NASA Exoplanets Research Program (\#80NSSC18K0349) and Habitable Worlds Program (\#80NSSC20K0226).  M.S.M. acknowledges support from GSFC Sellers Exoplanet En- vironments Collaboration (SEEC), with funding specifically by the NASA Astrophysics Divisions Internal Scientist Funding Model. This work was made possible by support from the UCSC Other Worlds Laboratory and the WFIRST Science Investigation Team program. The results reported herein benefited from
collaborations and/or information exchange within NASAÕs Nexus for Exoplanet System Science (NExSS) research coordination network, sponsored by NASAÕs Science Mission Directorate.  Computation for this research was performed by the UCSC Hyades supercomputer, which was supported by the National Science Foundation (award number AST-1229745).

\vspace{5mm}


\software{
    PyMultiNest \citep{buchner2014x}
    }












\end{document}